\documentstyle[12pt]{article}
\begin{document}
\def\ds{\displaystyle}
\def\beq{\begin{equation}}
\def\eeq{\end{equation}}
\def\bea{\begin{eqnarray}}
\def\eea{\end{eqnarray}}
\def\ve{\vert}
\def\vel{\left|}
\def\ver{\right|}
\def\nnb{\nonumber}
\def\ga{\left(}
\def\dr{\right)}
\def\aga{\left\{}
\def\adr{\right\}}
\def\lla{\left<}
\def\rra{\right>}
\def\rar{\rightarrow}
\def\nnb{\nonumber}
\def\la{\langle}
\def\ra{\rangle}
\def\ba{\begin{array}}
\def\ea{\end{array}}
\def\tr{\mbox{Tr}}
\def\ssp{{\Sigma^{*+}}}
\def\sso{{\Sigma^{*0}}}
\def\ssm{{\Sigma^{*-}}}
\def\xis0{{\Xi^{*0}}}
\def\xism{{\Xi^{*-}}}
\def\qs{\la \bar s s \ra}
\def\qu{\la \bar u u \ra}
\def\qd{\la \bar d d \ra}
\def\qq{\la \bar q q \ra}
\def\gGgG{\la g^2 G^2 \ra}
\def\q{\gamma_5 \not\!q}
\def\x{\gamma_5 \not\!x}
\def\g5{\gamma_5}
\def\sb{S_Q^{cf}}
\def\sd{S_d^{be}}
\def\su{S_u^{ad}}
\def\ss{S_s^{??}}
\def\sbp{{S}_Q^{'cf}}
\def\sdp{{S}_d^{'be}}
\def\sup{{S}_u^{'ad}}
\def\ssp{{S}_s^{'??}}
\def\sig{\sigma_{\mu \nu} \gamma_5 p^\mu q^\nu}
\def\fo{f_0(\frac{s_0}{M^2})}
\def\ffi{f_1(\frac{s_0}{M^2})}
\def\fii{f_2(\frac{s_0}{M^2})}
\def\O{{\cal O}}
\def\sl{{\Sigma^0 \Lambda}}
\def\es{\!\!\! &=& \!\!\!}
\def\ar{&+& \!\!\!}
\def\ek{&-& \!\!\!}
\def\cp{&\times& \!\!\!}
\def\se{\!\!\! &\simeq& \!\!\!}
\title{
         {\Large
                 {\bf
Radiative $\phi \rar f_0(980) \gamma$ decay in light cone QCD sum rules
                 }
         }
      }

\author{\vspace{1cm}\\
{\small T. M. Aliev$^a$ \thanks
{e-mail: taliev@metu.edu.tr}\,\,,
A. \"{O}zpineci$^b$ \thanks
{e-mail: ozpineci@ictp.trieste.it}\,\,,
M. Savc{\i}$^a$ \thanks
{e-mail: savci@metu.edu.tr}} \\
{\small a Physics Department, Middle East Technical University, 
06531 Ankara, Turkey}\\
{\small b  The Abdus Salam International Centre for Theoretical Physics,
I-34100, Trieste, Italy} }
\date{}

\begin{titlepage}
\maketitle
\thispagestyle{empty}

\begin{abstract}

The light cone QCD sum rules method is used to calculate the transition form
factor for the radiative $\phi \rar f_0 \gamma$ decay, assuming that
the quark content of the $f_0$ meson is pure $\bar s s$ state. 
The branching ratio is estimated to be 
${\cal B} (\phi \rar f_0 \gamma) = 3.5 \times (1 \pm 0.3)\times 10^{-4}$. 
A comparison of  our prediction on branching ratio with the theoretical 
results and experimental data existing in literature is presented.  

\end{abstract}

~~~PACS number(s): 11.55.Hx, 13.40.Hq, 14.40.Ev
\end{titlepage}

\section{Introduction}
According to the quark model, mesons are interpreted as pure $\bar q
q$ states. Scalar mesons constitute a remarkable exception to this
systematization and their nature is not well established yet
\cite{R1}--\cite{R4}.

In the naive $\bar q q$ picture, one can treat the isoscalar $f_0(980)$
either as the meson that exists mostly as nonstrange and almost
degenerate with the isovector $a_0(980)$ or as mainly
$\bar s s$, in analogy to the pure $\bar s s$ vector meson $\phi(1020)$.

In order to understand the content of the $f_0$ meson several alternatives
have been suggested, such as, the analysis of the $f_0 \rar 2 \gamma$ decay
\cite{R5,R6}; study of the ratio $\Gamma(a_0 \rar f_0 \gamma) /
\Gamma (\phi \rar f_0 \gamma)$ \cite{R7}. However, among these, the $(\phi
\rar f_0 \gamma)$ decay occupies a special place, since the branching ratio
expected of this decay, is essentially dependent on the content of $f_0$.
For example ${\cal B}(\phi \rar f_0 \gamma)$ is as high as $\sim 10^{-4}$ if
it were composed of $\bar q q \bar q q $, and $\sim 10^{-5}$ if $f_0$ were a
pure $\bar s s$ state.

It has been known for a long
time that $f_0(980)$ couples significantly through its $\bar s s$ content,
from its detection as a peak in the $J/\psi \rar \phi f_0$ 
\cite{R8} and $D_s \rar \pi f_0$ \cite{R9} decays, 
as discussed in \cite{R10} and \cite{R11} (see also \cite{R12}).
For this reason, in this work we assume that quark content of both $\phi$ 
and $f_0$ mesons are pure $\bar s s$.   
In the present paper we analyze the radiative $\phi \rar f_0 \gamma$ decay
in framework of the light cone QCD sum rules (about light cone QCD sum rules
and its applications, see for example \cite{R13}).
Note also that the $\phi \rar f_0 \gamma$ decay is analyzed in framework of
the 3--point sum rules in \cite{R14}.
In order to calculate the transition form factor describing the $\phi \rar
f_0 \gamma$ decay in light cone QCD sum rules, we consider the following 
correlator
\bea
\label{e1}
\Pi_\mu = i \int d^4 x e^{ipx} \lla 0 \vel T\Big\{
J^s(x) J^\phi_\mu (0)\Big\}\ver 0 \rra_\gamma~,
\eea   
where $J^s = \bar s s$ and $J^\phi_\mu= \bar s \gamma_\mu s$  
are interpolating currents for $f_0$ and $\phi$ mesons, respectively, and
$\gamma$ is the background electromagnetic field (for more about external field
technique in QCD see \cite{R15,R16}).  

The physical part of the correlator can be obtained by inserting a complete
set of one meson states into the correlator,
\bea
\label{e2}
\Pi_\mu = \sum \frac{\lla 0 \ve J^s(x) \ve f_0(p) \rra \lla f_0(p)
\ve \phi(p_1) \rra_\gamma \lla \phi(p_1) \ve J_\mu^\phi (0)\ve 0 \rra}
{\ga p^2 - m_{f_0}^2\dr \ga p_1^2 - m_\phi^2 \dr}~,
\eea
where $\phi$ and $f_0$ are the quantum numbers and $p_1=p+q$ with $q$ being the
photon momentum. 

The matrix element $\lla \phi(p_1) \vel J_\mu^\phi(0) \ver 0 \rra$ in Eq.
(\ref{e1}) is defined as
\bea
\label{e3}
\lla \phi(p_1) \vel J_\mu^\phi(0) \ver 0 \rra = m_\phi f_\phi
\varepsilon_\mu^\phi~,
\eea
where $\varepsilon_\mu^\phi$ is the $\phi$ meson polarization vector.
The coupling of the $f_0(980)$ to the
scalar current $J^s = \bar s s$ is defined in terms of a constant
$\lambda_f$
\bea
\label{e4}
\lla 0 \vel J^s \ver f_0(p) \rra = m_{f_0} \lambda_f~.
\eea

The relevant matrix element describing the transition $\phi \rar f_0$
induced by an external electromagnetic current can be parametrized in the
following form:
\bea
\label{e5}
\lla f_0(p) \ve \phi(p_1,\varepsilon^\phi) \rra_\gamma = e \varepsilon^\mu
\Big[ F_1 (q^2) (p_1 q) \varepsilon^\phi_\mu + F_2 (q^2) (\varepsilon^\phi
q) p_{1\mu} \Big]~,
\eea
where $\varepsilon$ is the photon polarization and we have used
$(\varepsilon q)=0$. From gauge invariance we have
\bea
\label{e6}
F_1(q^2) = - F_2(q^2),
\eea
and since the photon is real in the decay under consideration, we need the
values of the form factors only at the point $q^2=0$. Using Eq. (\ref{e6})
the matrix element
$\lla f_0 \ve \phi \rra_\gamma$ takes the following gauge invariant form,
\bea
\label{e7}
\lla f_0 \ve \phi \rra_\gamma = e \varepsilon^\mu F_1(0) \Big[ 
(p_1 q) \varepsilon^\phi_\mu - (\varepsilon^\phi q) p_{1\mu} \Big]~.
\eea
Using Eqs. (\ref{e1})--(\ref{e4}) and (\ref{e7}), for the
phenomenological part of the correlator we have
\bea
\label{e8}
\Gamma_\mu^{phen} = e F_1(0) \varepsilon^\nu \Big[-(p_1 q) g_{\mu\nu} +
p_{1\nu} q_\mu \Big] \frac{\lambda_f f_\phi m_{f_0} m_\phi}
{\ga p^2-m_{f_0}^2\dr \ga  p_1^2-m_\phi^2\dr}~.
\eea
In order to construct the sum rule, calculation of the correlator 
from QCD side (theoretical part) is needed. From Eq. (\ref{e1}) we get
\bea
\label{e9}
\Pi_\mu = \int d^4x e^{ipx} \lla 0 \vel \mbox{\rm Tr} \Big\{
-\gamma_\mu S_s(-x) S_s(x) \Big\} \ver 0 \rra_\gamma~,
\eea
where $S_s$ is the full propagator of the strange quark (see below). Theoretical
part of the correlator contains two pieces,  perturbative and nonperturbative. 
Perturbative part corresponds to the case when photon is radiated from the
freely propagating quarks. Its expression can be obtained by making the
following replacement in each one of the quark propagators in Eq. (\ref{e9})
\bea
\label{e10}
{S_s}_{\alpha\beta}^{ab} \rar 2 e e_q \Big( dy F_{\mu\nu} y^\nu S_s^{free}(x-y)
\gamma^\mu S_s^{free} (y) \Big)_{\alpha\beta}^{ab}~,
\eea
where the Fock--Schwinger gauge $x^\mu A_\mu(x)=0$ is used and $S_s^{free}$ 
is the free s--quark propagator $S_s^{free}(x)=i\not\! x/(2 \pi^2 x^4)$
and the remaining one is the full quark propagator.  

The nonperturbative piece of the theoretical part can be obtained 
from Eq. (\ref{e9}) by replacing each one of the propagators with
\bea
\label{e11} 
{S_s}_{\alpha\beta}^{ab} = -\frac{1}{4} \bar q^a A_i q^b 
(A_i)_{\alpha\beta}~,
\eea
where $A_i$ is the full set of Dirac matrices and sum over $A_i$ is implied
and the other quark propagator is the full propagator, involving pertubative
and nonperturbative contributions. In order to calculate perturbative and
nonperturbative parts to the correlator function (\ref{e1}),
expression of the s--quark propagator in external field is needed. 

The complete light cone expansion of the light quark propagator in external
field is presented in \cite{R16}. The propagator receives contributions from
the nonlocal operators $\bar q G q$, $\bar q G G q$, $\bar q q \bar q q$,
where G is the gluon field strength tensor. In the present work  
we consider operators with only one gluon
field and neglect terms with two gluons $\bar q GG q$, and four
quarks $\bar q q \bar q q$ and formal neglect of these these terms
can be justified on the basis of an expansion in
conformal spin \cite{R17}.  In this approximation full propagator of the 
s--quark is given as
\bea
\label{e12}
S_s(x) \es \frac{i \not\!x}{2 \pi^2 x^4} - \frac{\qs}{12} \ga 1+
\frac{x^2}{16} m_0^2 \dr + \frac{i m_s \qs}{48} \not\!x -
\frac{i m_0^2 m_s}{2^7 3^2} x^2 \not\!x \nnb \\
\ek ig_s \int_0^1 
dv \left[\frac{\not\!x}{16\pi^2 x^2} G_{\mu\nu}(vx)\sigma^{\mu\nu} -
\frac{i}{4 \pi^2 x^2} v x^\mu G_{\mu\nu}\gamma^\nu \right]~.
\eea 
It follows from Eqs. (\ref{e11}) and (\ref{e9}) that in calculating the QCD
part of the correlator, as is generally the case, we are left with the matrix 
elements of the gauge invariant nonlocal operators, sandwiched in between 
the photon and the vacuum states 
$\lla \gamma(q) \vel \bar s A_i s \ver 0 \rra$. These matrix 
elements define the
light cone photon wave functions. The photon wave functions up to twist--4
are \cite{R17,R18}
\bea
\label{e13}
\lla \gamma(q) \vel \bar q(x) \sigma_{\mu\nu} q(0) \ver 0 \rra \es
i e e_q \qq \int_0^1 du\, e^{iqx} \Big\{\ga \varepsilon_\mu q_\nu -
\varepsilon_\nu q_\mu \dr\Big[
\chi \phi(u) + x^2 \Big( g_1(u) - g_2(u) \Big) \Big] \nnb \\
\ar \Big[ (qx)
(\varepsilon_\mu x_\nu - \varepsilon_\nu x_\mu) + (\varepsilon x) 
(x_\mu q_\nu - x_\nu q_\mu) \Big] g_2(u) \Big\}~, \\ \nnb \\
\label{e14}
\lla \gamma(q) \vel \bar q(x) \gamma_\mu \gamma_5 q(0) \ver 0 \rra \es
\frac{e f}{4} e_q \epsilon_{\alpha\beta\rho\sigma} \varepsilon^\beta q^\rho
x^\sigma \int_0^1 du \,e^{iuqx} \psi(u)~.
\eea 
The path--ordered gauge factor ${\cal P} exp \Big( i g_s \int_0^1 du\, x^\mu
A_\mu (ux) \Big)$ is emitted since the Schwinger--Fock gauge $x^\mu A_\mu(x)
= 0$ is used. The functions $\phi(u),~\psi(u)$ are the leading twist--2
photon wave functions, while $g_1(u)$ and $g_2(u)$ are the twist--4 photon
wave functions. Note that twist--3 photon wave functions are neglected in
the calculations, since their contributions are small and change the result
by 5\%. In Eq. (\ref{e13}) $\chi$ is the magnetic
susceptibility of the quark condensate and $e_q$ is the quark charge. The
theoretical part is obtained by substituting photon wave functions and
expression for the $s$--quark propagators into Eq. (\ref{e9}). The sum rules
is obtained by equating the phenomenological and theoretical parts of the
correlator. In order to suppress higher states and continuum contribution
(for more details see \cite{R19,R20}) double Borel transformations of the
variables $p_1^2=p^2$ and $p_2^2=(p+q)^2$ are performed on both sides of the
correlator, after which the following sum rule is obtained
\bea
\label{e15}
F_1(0) \es e^{m_{f_0}^2/M_2^2} e^{m_{\phi}^2/M_1^2} \frac{e_s}{\lambda_{f_0}
f_\phi m_{f_0} m_\phi}\Bigg\{ \Bigg[ 2 \chi \qs \phi(u_0) - \frac{3 m_s}{2
\pi^2} (1 + \gamma_E) \Bigg] M^2 E_0(s_0^2/M^2) \nnb \\
\ar \frac{1}{24} \qs \Big[ -192 g_1(u_0) + m_s \phi(u_0) \qs \Big] \nnb \\
\ar \frac{3 m_s}{2 \pi^2} \Bigg[ M^2 \Bigg( \gamma_E + \mbox{\rm ln}
\frac{M^2}{\Lambda^2} \Bigg)  E_0 (s_0/M^2) + M^2 f(s_0/M^2) \Bigg]
\Bigg\}~,
\eea
where $s_0$ is the continuum threshold 
\bea
E_0(s_0/M^2) = 1 - e^{-s_0/M^2}~, \nnb \\
f(s_0/M^2) = \int_0^{s_0/M^2} dy \, \mbox{\rm ln}y \, e^{-y}~,\nnb
\eea
which have been used to subtract continuum, and
\bea
u_0 = \frac{M_2^2}{M_1^2+M_2^2}~,~~~~~
M^2 = \frac{M_1^2 M_2^2}{M_1^2+M_2^2}~,\nnb
\eea
where $M_1^2$ and $M_2^2$ are the Borel parameters in $\phi$ and $f_0$
channels, respectively, 
$\Lambda$ is the QCD scale parameter and $\gamma_E$ is the Euler
constant. Since the masses of $\phi$ and $f_0$ are very close to each other
we will set $M_1^2=M_2^2\equiv 2 M^2$, obviously from which it follows that
$u_0=1/2$.      

It is clear from Eq. (\ref{e14}) that the values of $\lambda_{f_0}$ and
$f_\phi$ are needed in order to determine $F(0)$. The coupling of the
$f_0(980)$ to the scalar $\bar s s$ current is determined by the constant
$\lambda_{f_0}$ and in the two--point QCD sum rules its value is found to be
$\lambda_{f_0} = (0.18 \pm 0.0015)~GeV$ \cite{R14}. 
In further numerical analysis we
will use $f_\phi=0.234~GeV$ which is obtained from the experimental analysis
of the $\phi \rar e^+ e^-$ decay \cite{R21}. 

Having the values of $\lambda_{f_0}$ and $f_\phi$, our next and final attempt
is the calculation of transition form factor $F_1(0)$.
As we can easily see from Eq. (\ref{e15}) the main input parameters of the
light cone QCD sum rules is the photon wave function. It is known that the
leading photon wave function receive only small corrections from the higher
conformal spin \cite{R17,R19,R22}, so that they do not deviate much from the
asymptotic form. The photon wave functions we use in our numerical analysis
are given as
\bea
\phi(u) = 6 u (1-u) ~, \nnb \\
\psi(u) = 1 ~, \nnb \\
g_1(u)  = - \frac{1}{8} (1-u) (3-u) ~. \nnb
\eea
Furthermore, the values of the input parameters that are used in the
numerical calculations are: $f=0.028~GeV^2,~\chi=-4.4~GeV^{-2}
$ \cite{R23} (in  \cite{R24} this
quantity is predicted to have the value $\chi=-3.3~GeV^{-2}$), 
$\lla \bar s s(1~GeV)\rra = -0.8\times (0.243)^3~GeV^3$ and the 
QCD scale parameter
is taken as $\Lambda = 0.2~GeV$. The strange quark mass is chosen in the
range $m_s = 0.125 \div 0.16~GeV$, obtained in the QCD sum rules approach
\cite{R25}. The masses of the $\phi$ and $f_0$ mesons are
$m_\phi=1.02~GeV,~m_{f_0}=0.98~GeV$. The transition form factor is a
physical quantity and therefore it must be independent of the auxiliary
continuum threshold $s_0$ and and the Borel mass $M^2$ parameters. So our
main concern is to find a region where the transition form factor $F_1(0)$
is practically independent of the parameters $s_0$ and $M^2$. For this aim
in Fig. (1) we present the dependence
of the transition form factor $F_1(0)$ on the Borel parameter $M^2$ at three
different values of the continuum threshold:
$s_0=2.0~GeV^2,~2.2~GeV^2$ and $2.4~GeV^2$. It follows from this
figure that for the choice of the continuum thresholds in the
above--mentioned range, the variation of the result on the transition form
factor $F_1(0)$ is about $10\%$. In other words, we can conclude that
$F_1(0)$ is practically independent of the continuum threshold. Furthermore
we observe that when $1.4 \le M^2 \le 2.0~GeV^2$, $F_1(0)$ is quite stable
with respect to the variations of the Borel parameter $M^2$. As a result, one
can directly read from this figure 

\bea
F_1(0) = (3.25 \pm 0.20)~GeV^{-1}~,\nnb
\eea
where the resulting error is due to the variations in $s_0$ and $M^2$. The
other sources of errors contributing to the numerical analysis of the 
transition form factor come from the strange quark mass and the
uncertainties in values of various condensates. Hence, our final prediction on the
transition form factor is
\bea
\label{e16}
F_1(0) = (3.25 \pm 0.50)~GeV^{-1}~.
\eea

Using the matrix element (\ref{e7}) for the decay width of the considered
process, we obtain
\bea
\label{e17}
\Gamma(\phi \rar f_0\gamma) = \alpha \vel F_1(0) \ver^2
\frac{\ga m_\phi^2-m_{f_0}^2\dr^3}{24 m_\phi^3}~.
\eea
Using the experimental value $\Gamma_{tot}(\phi)=4.458~MeV$
\cite{R21}, and Eqs. (\ref{e16}) and (\ref{e17}), we get for the branching ratio
\bea
\label{e18}
{\cal B}(\phi \rar f_0\gamma) = 3.5\times (1.0\pm0.3)\times 10^{-4}~.
\eea

Our result on the branching ratio is obtained under the assumption that
$f_0$ meson is represented as a pure $\bar s s$ component.
How does the result change if we assume that $\phi$ and $f_0$ mesons can be
represented as a mixing of $\bar s s$ and $\bar n n=(\bar u u+\bar d d)/\sqrt{2}$ 
state, i.e.,
\bea
\phi \es \cos\alpha \, \bar s s + \sin\alpha \, \bar n n~,\nnb\\
f_0 \es \sin\beta \, \bar s s + \cos\beta \, \bar n n~?\nnb
\eea
Analysis of the process $\phi \rar \pi^0 \gamma$ and combined analysis of
the $\phi \rar f_0 \gamma$ and $f_0 \rar 2\gamma$ decays show that $\vel
\alpha \ver \le 4^0$ and two solutions are found for $\beta$, i.e., 
$\beta = -48^0 \pm 6^0$ or $\beta = 85^0 \pm 5^0$, respectively (see for
example \cite{R5}).  In other words, quark content of $\phi$ meson is pure 
$\bar s s$ state, while in $f_0$ meson there might be sizable $\bar n n$
component. Obviously, when $F_1(0)$ is calculated from QCD side, only
$\sin\beta \, \bar s s$ component operates (see Eq. (\ref{e1})) 
and therefore the decay width
$\Gamma(\phi \rar f_0\gamma)$, and hence the corresponding branching ratio,
contains an extra factor $\sin^2\beta$.
If $\beta = 85^0 \pm 5^0$, then prediction for the branching ratio given in
Eq. (\ref{e18}) is practically unchanged, but when $\beta = -48^0 \pm
6^0$ ${\cal B}(\phi \rar f_0 \gamma)$ decreases by about a factor of 2.

Finally let us compare our prediction on branching ratio with the existing
theoretical results and experimental data in the literature. Obviously, our
result is slightly larger compare to the 3--point QCD sum rule result 
which predicts 
${\cal B}(\phi \rar f_0 \gamma) \simeq (2.7 \pm 1.1)\times 10^{-4}$
\cite{R14}, and approximately three times 
larger compared to the prediction of the spectral QCD sum
rules and chiral unitary approaches, whose predictions are 
${\cal B}(\phi \rar f_0 \gamma)=1.3 \times 10^{-4}$ \cite{R26} and
${\cal B}(\phi \rar f_0 \gamma)=1.6 \times 10^{-4}$ \cite{R27},
respectively.
It is interesting to note that this value of the branching ratio
is closer to our prediction when the mixing angle is chosen to be
$\beta=-48^0\pm 6^0$. Our result, which is given in Eq. (\ref{e18}), is larger
compared to the predictions of \cite{R28,R7}, whose results are 
${\cal B}(\phi \rar f_0 \gamma)= 1.9\times 10^{-4}$ \cite{R28}, and
${\cal B}(\phi \rar f_0 \gamma)= 1.35\times 10^{-4}$ \cite{R7},
respectively.

As the final words we would like to point out that our prediction given in
Eq. (\ref{e18}), is in a very good agreement with the existing experimental 
result ${\cal B}(\phi \rar f_0 \gamma) \simeq (3.4 \pm 1.1)\times 10^{-4}$
\cite{R21}.

\newpage

\begin{figure}
\vskip 1.5 cm
    \includegraphics{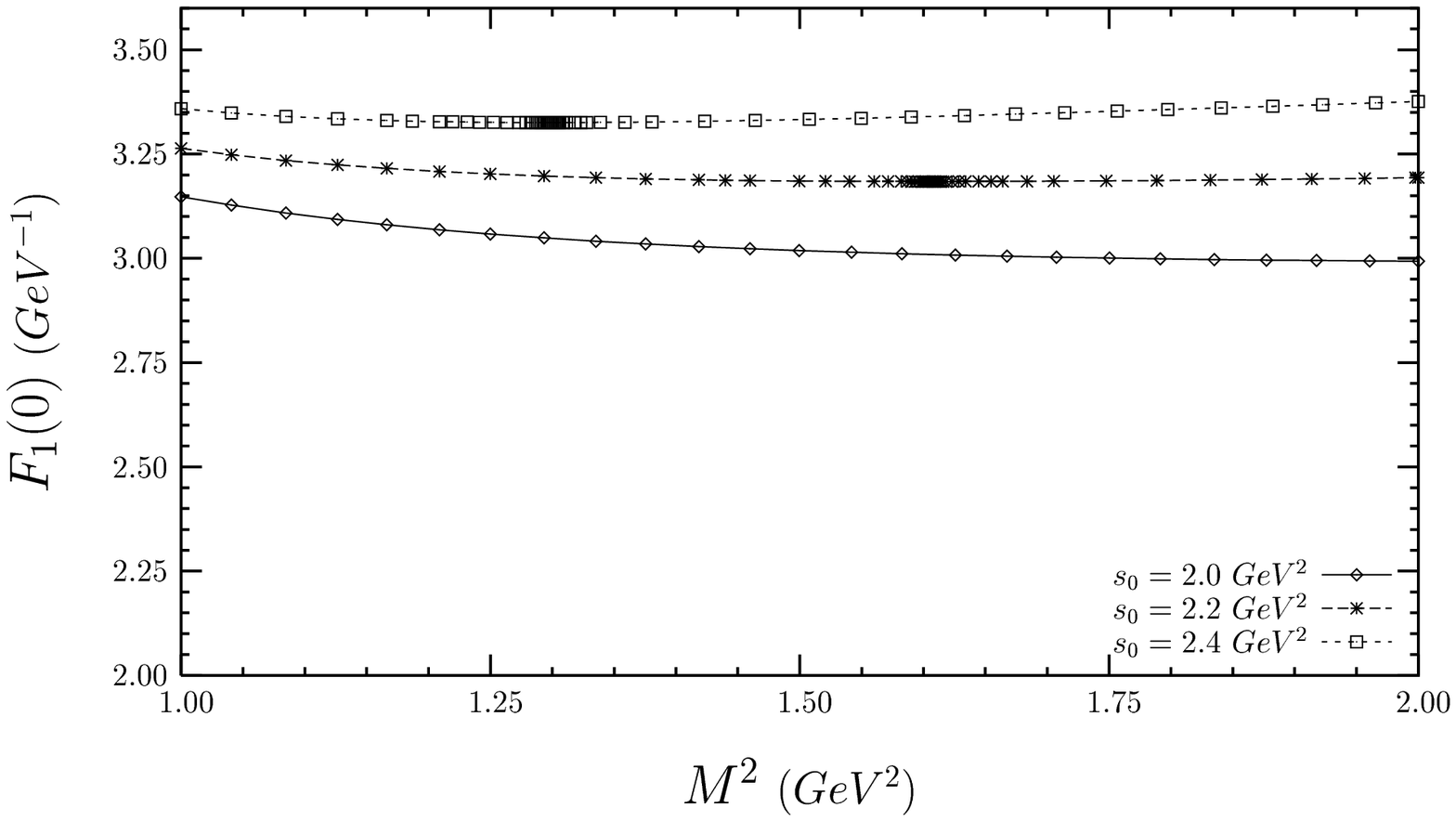}
\vskip 7.8cm
\caption{The dependence of the transition form factor $F_1(0)$ for the
radiative $\phi \rar f_0 \gamma$ decay on
$M^2$ at three different values of the
continuum threshold $s_0=2.0~GeV^2,~2.2~GeV^2$ and $2.4~GeV^2$. }
\end{figure}  


\newpage
\begin{thebibliography}{99}


\bibitem{R1} L. Montanet,
{\it Rep. Prog. Phys.} {\bf 46} (1983) 337;\\
F. E. Close,
{\it Rep. Prog. Phys.} {\bf 51} (1988) 833;\\
N. N. Achasov,
{\it Nucl. Phys. Proc. Suppl.} {\bf B21} (1991) 189;\\
T. Barnes,
Prep. hep--ph/0001326 (2000);\\
V. V. Anisovich, 
Prep. hep--ph/0110326 (2001).

\bibitem{R2} N. A. Tornqvist,
{\it Phys. Rev. Lett.} {\bf 49} (1982) 624;
{\it Z. Phys.} {\bf C68} (1995) 647.

\bibitem{R3} N. A. Tornqvist and M. Ross,
{\it Phys. Rev. Lett.} {\bf 76} (1996) 1575.

\bibitem{R4} E. van Beveren {\it et al.},
{\it Z. Phys.} {\bf C30} (1986) 615;\\
E. van Beveren, G. Rupp and M. D. Scadron,
{\it Phys. Lett.} {\bf B495} (2000) 300;\\
ibid, {\bf B509} (2001) 365 (erratum).

\bibitem{R5} A. V. Anisovich, V. V. Anisovich and V. A. Nikonov,
Prep. hep--ph/0011191 (2000).

\bibitem{R6} M. Boglione and M. R. Pennington,
{\it Phys. Rev. Lett.} {\bf 79} (1997) 1998.

\bibitem{R7} F. E. Close, N. Isgur and S. Kumano,
{\it Nucl. Phys.} {\bf B389} (1993) 513;\\
N. Brown and F. E. Close,
in proc. "The DAFNE Physics Handbook", Eds. L. Maiani, G. Pancheri and N.
Paver, INFN, Frascati, 1995.

\bibitem{R8} G. Gidal {\it et al.}, MARK II Collab.,
{\it Phys. Lett.} {\bf B107} (1981) 153;\\
A. Falvard {\it et al.}, DM2 Collab.,  
{\it Phys. Rev.} {\bf D38} (1988) 2706.

\bibitem{R9} J. C. Anjos {\it et al.}, E691 Collab.,      
{\it Phys. Rev. Lett.} {\bf 62} (1989) 125;\\
E. M. Aitala {\it et al.}, E791 Collab.,
{\it Phys. Rev. Lett.} {\bf 86} (2001) 125.

\bibitem{R10} K. L. Au, D. Morgan and M. R. Pennington,
{\it Phys. Rev.} {\bf D35} (1987) 1633.

\bibitem{R11} D. Morgan and M. R. Pennington,
{\it Phys. Rev.} {\bf D48} (1993) 1185.

\bibitem{R12} R. Delbourgo, D. --S. Liu and M. D. Scadron,
{\it Phys. Lett.} {\bf B446} (1999) 332.

\bibitem{R13} P. Colangelo and A. Khodjamirian,
In, "At the Frontier of Particle
Physics, Handbook of QCD", ed. by M. Shifman, World Scientific, 
Singapore, (2001) 1495.    

\bibitem{R14} F. De Fazio and M. R. Pennington,
{\it Phys. Lett.} {\bf B520} (2001) 78. 

\bibitem{R15} B. L. Ioffe, A. V. Smilga,
{\it Nucl. Phys.} {\bf B232} (1984) 109.

\bibitem{R16} I. I. Balitsky and V. M. Braun,
{\it Nucl. Phys.} {\bf B311} (1988) 541. 

\bibitem{R17} V. M. Braun and I. E. Filyanov,
{\it Z. Phys.} {\bf C48} (1990) 239.

\bibitem{R18} A. Ali, V. M. Braun,  
{\it Phys. Lett.} {\bf B359} (1995) 223. 

\bibitem{R19} V. M. Belyaev, V. M. Braun, A. Khodjamirian and R.
R{\"u}ckl,\\
{\it Phys. Rev.} {\bf D51} (1995) 6177.

\bibitem{R20} T. M. Aliev, A. \"{O}zpineci, M. Savc{\i},
{\it Nucl. Phys.} {\bf A678} (2000) 443;\\
{\it Phys. Rev.} {\bf D62} (2000) 053012.

\bibitem{R21} Particle Data Group, D. E. Groom, {\it et al.},
{\it Eur. Phys. J.} {\bf C15} (2000) 1.

\bibitem{R22} I. I. Balitsky, V. M. Braun, A. V. Kolesnichenko,
{\it Nucl. Phys.} {\bf B312} (1989) 509;\\
V. M. Braun and I. E. Filyanov,
{\it Z. Phys.} {\bf C44} (1989) 157.
 
\bibitem{R23} V. M. Belyaev, Ya. I. Kogan,
{\it Yad. Fiz.} {\bf 40} (1984) 1035
({\it Sov. J. Nucl. Phys.} {\bf 40} (1984) 659).   

\bibitem{R24} I. I. Balitsky and A. V. Kolesnichenko,
{\it Yad. Fiz.} {\bf 41} (1985) 282
({\it Sov. J. Nucl. Phys.} {\bf 41} (1985) 178).  

\bibitem{R25} P. Colangelo, F. De Fazio, G. Nardulli and N. Paver,
{\it Phys. Lett.} {\bf B408} (1997) 340.

\bibitem{R26} S. Narison,
{\it Nucl. Phys. Proc. Suppl.} {\bf 96} (2001) 244.

\bibitem{R27} E. Marco, S. Hirenzaki, E. Oset and H. Toki,
{\it Phys. Lett.} {\bf B470} (1999) 20. 

\bibitem{R28} J. Lucio and J. Pestiean,
{\it Phys. Rev.} {\bf D42} (1990) 3253.


\end{thebibliography}
\end{document}